\documentclass[twocolumn]{aastex62}

\usepackage{amsmath,amssymb,graphicx}
\hypersetup{breaklinks=true,colorlinks=true,citecolor=blue,linkcolor=blue,urlcolor=blue}

\shorttitle{SN Ia late-time near-IR light curves}
\shortauthors{Graur et al.}

\begin{document}

\title{A year-long plateau in the late-time near-infrared light curves of Type Ia supernovae}

\correspondingauthor{Or Graur}
\email{or.graur@cfa.harvard.edu}

\author{Or Graur}
\affiliation{Harvard-Smithsonian Center for Astrophysics, 60 Garden St., Cambridge, MA 02138, USA}
\affiliation{Department of Astrophysics, American Museum of Natural History, New York, NY 10024, USA}
\affiliation{NSF Astronomy and Astrophysics Postdoctoral Fellow}

\author{Kate Maguire}
\affiliation{School of Physics, Trinity College Dublin, Dublin 2, Ireland}

\author{Russell Ryan}
\affiliation{Space Telescope Science Institute, Baltimore, MD 21218, USA}

\author{Matt Nicholl}
\affiliation{Institute for Astronomy, University of Edinburgh, Royal Observatory, Blackford Hill, EH9 3HJ, UK}
\affiliation{Birmingham Institute for Gravitational Wave Astronomy and School of Physics and Astronomy, University of Birmingham, Birmingham B15 2TT, UK}

\author{Arturo Avelino}
\affiliation{Harvard-Smithsonian Center for Astrophysics, 60 Garden St., Cambridge, MA 02138, USA}

\author{Adam G. Riess}
\affiliation{Space Telescope Science Institute, Baltimore, MD 21218, USA}
\affiliation{Department of Physics and Astronomy, The Johns Hopkins University, Baltimore, MD 21218, USA}

\author{Luke Shingles}
\affiliation{Astrophysics Research Centre, School of Mathematics and Physics, Queen’s University Belfast, Belfast BT7 1NN, UK}

\author{Ivo R. Seitenzahl}
\affiliation{School of Science, University of New South Wales, Australian Defense Force Academy, Canberra, ACT 2600, Australia}

\author{Robert Fisher}
\affiliation{Department of Physics, University of Massachusetts Dartmouth, 285 Old Westport Road, North Dartmouth, MA 02740, USA}


\begin{abstract}
 The light curves of Type Ia supernovae are routinely used to constrain cosmology models. Driven by radioactive decay of $^{56}$Ni, the light curves steadily decline over time, but $>150$ days past explosion, the near-infrared portion is poorly characterized. We report a year-long plateau in the near-infrared light curve at 150--500 days, followed by a second decline phase accompanied by a possible appearance of [Fe I] emission lines. This near-infrared plateau contrasts sharply with Type IIP plateaus and requires a new physical mechanism. We suggest a such as masking of the ``near-infrared catastrophe,'' a predicted yet unobserved sharp light-curve decline, by scattering of ultraviolet photons to longer wavelengths. The transition off the plateau could be due to a change in the dominant ionization state of the supernova ejecta. Our results shed new light on the complex radiative transfer processes that take place in Type Ia supernovae and enhance their use as ``standard candles.''
\end{abstract}



\section{Introduction}
\label{sec:intro}

The optical and near-infrared (NIR) light we receive from Type Ia supernovae (SNe Ia) at early times is a product of the radioactive decay chain $^{56}$Ni$\to^{56}$Co$\to^{56}$Fe. This decay chain proceeds through inverse $\beta$ decay, i.e., electron capture or positron emission. At each step, the daughter nucleus de-excites down to the ground state by emitting high-energy $\gamma$ rays. Together with positrons emitted by the inverse $\beta$ decay process, these $\gamma$ rays are then thermalized by the expanding SN ejecta and downconverted to optical and NIR photons \citep{Truran1967,Colgate1969,1982ApJ...253..785A}. 

\begin{figure*}
 \centering
 \includegraphics[width=\textwidth]{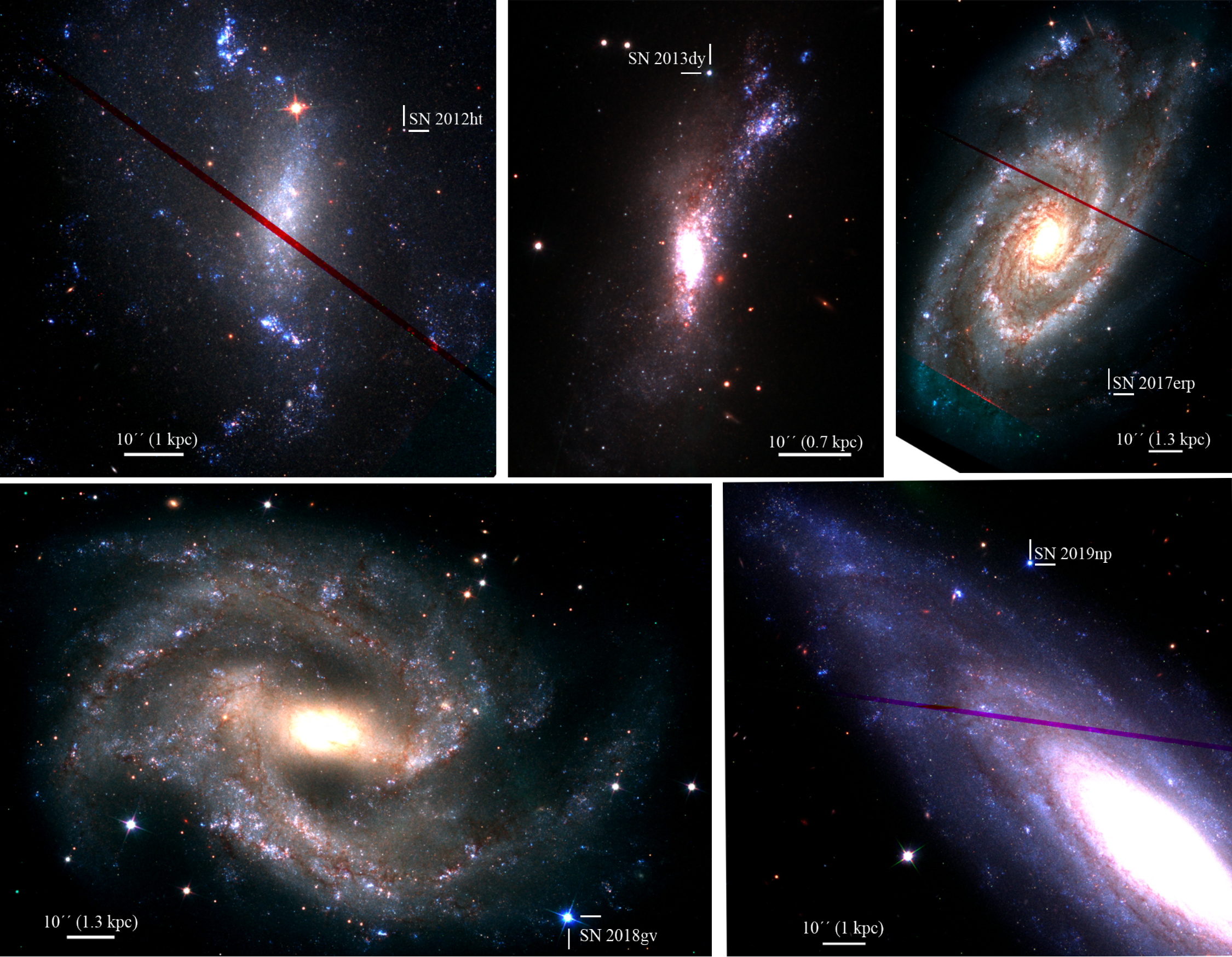}
 \caption{\textit{HST} color composites of SNe 2012ht, 2013dy, 2017erp, 2018gv, and 2019np in their host galaxies. The color images are composed of \textit{F160W} (red), \textit{F814W} (green), and \textit{F350LP} (blue) images from all visits to the host galaxies. In each image, the location of the SN is marked by a by a white reticle. North is up and East is to the left.}
 \label{fig:hosts}
\end{figure*}

This picture was known to hold out to $\sim 500$ days past maximum light, at which point it was predicted that the cooling of the ejecta would transition from collisionally-excited emission lines in the optical and NIR to fine-structure iron lines in the mid and far IR \citep{1980PhDT.........1A}. This transition was supposed to manifest in a sharp decline in the optical and NIR light. To date, this so-called ``IR catastrophe'' has not been observed in the optical. In fact, several recent studies have shown that the decline of the optical light curves begins to \emph{slow down} at $\sim 800$ days \citep{1997A&A...328..203C,2009A&A...505..265L,2015ApJ...814L...2F,2016ApJ...819...31G,2018ApJ...859...79G,2018ApJ...866...10G,2019ApJ...870...14G,2017ApJ...841...48S,2018ApJ...852...89Y,2017MNRAS.468.3798D,2017MNRAS.472.2534K,2019ApJ...882...30L}.

Here, we report observations of five nearby SNe Ia (SNe 2012ht, 2013dy, 2017erp, 2018gv, and 2019np) with the \textit{Hubble Space Telescope} (\textit{HST}) Wide-Field Camera 3 (WFC3) in the optical \textit{F350LP}, \textit{F555W}, and \textit{F814W} filters and the NIR \textit{F160W} filter (presented in Supplementary Table~1). We also present NIR spectra of SN 2017erp, including an \textit{HST} grism spectrum at $>500$ days past $B$-band maximum light (from here on out, all SN phases will be reported in relation to $B$-band maximum light). Figure~1 shows composite color images of the SNe and their locations within their host galaxies. 

Instead of a steep drop in the NIR light curves, as predicted by the ``IR catastrophe,'' we report a year-long plateau in the $J$ and $H$ bands that sets in at $\approx150$ days and lasts until $\approx400$--$500$ days. After the plateau, the SNe transition into a new phase during which the $H$-band light curve declines at the same rate as the optical light curve, consistent with the continued radioactive decay of $^{56}$Co$\to^{56}$Fe. 

The occurrence of a plateau in the NIR light curve is intriguing, since to date only one type of SN---the hydrogen-rich SNe IIP---have been known to exhibit a plateau. In SNe IIP, the plateau sets in shortly after peak, lasts $\approx 100$ days, and is due to the cooling of the shock-heated ejecta and recombination of hydrogen \citep{1976ApJ...207..872C}. Since the vast majority of SNe Ia, by definition, lack hydrogen, the NIR plateau reported here must be the result of a different physical process.

\begin{figure*}
 \centering
 \includegraphics[width=\textwidth]{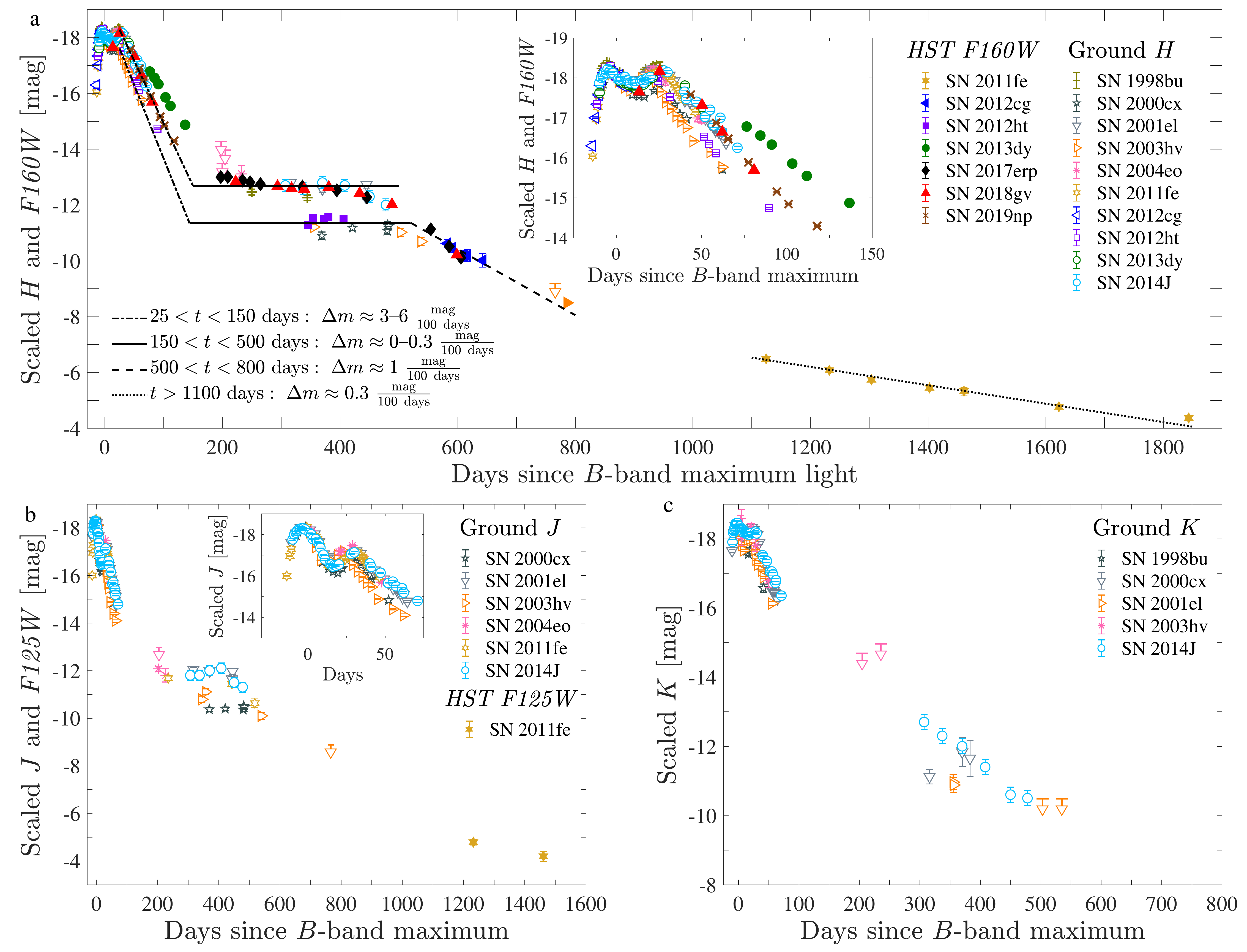}
 \caption{SN Ia near-infrared light curves. Although standard to $\sim 0.1$ mag at peak, the $H$-band light curves (\textbf{a}) begin to branch out after the second peak, with decline rates in the range 3--$6~{\rm mag/100~days}$. At $\approx 150$ days, the light curves settle into a plateau phase that lasts until $\approx 400$--$500$ days, when they once again transition into a second decline phase with a rate of $\approx 1~{\rm mag/100~days}$. At the plateau phase, the SNe have a range of $\sim 2$ mag, where the brighter SNe had slower decline rates before entering the plateau. The measurements in this plot have been scaled to the light curve of SN 2011fe. Using this scaling, the plateau phase is also apparent in the $J$ band (\textbf{b}). Synthetic photometry of SN 2014J in the $K$ band (\textbf{c}) show no hint of a plateau in this wavelength range. Black curves, meant to guide the eye, represent the distinct phases of the $H$-band light curve. Representative decline rates along each phase are noted in the legend at the bottom of the upper panel. Error bars represent $1\sigma$ measurement uncertainties, while downward arrows indicate $3\sigma$ upper limits (see refs.~\cite{2009A&A...505..265L,2007MNRAS.377.1531P}). The data behind this figure are provided in Supplementary Data 1.}
 \label{fig:lc}
\end{figure*}


\section{Results}
\label{sec:results}

In Figure~2, we combine our \textit{F160W} measurements with \textit{HST} \textit{F160W} and ground-based $H$-band observations of the SNe in our sample as well as several SNe Ia from the literature \citep{2009A&A...505..265L,2016ApJ...819...31G,2017ApJ...841...48S,2003PASP..115..277C,2003AJ....125..166K,2004A&A...428..555S,2007A&A...470L...1S,2007MNRAS.377.1531P,2012ApJ...754...19M,2015MNRAS.452.4307P,2016ApJ...820...92M,2016ApJ...822L..16S,2018ApJ...869...56B}. We also show literature $J$- and $K$-band data, where available, for the same SNe. Finally, we add synthetic photometry of SN 2014J, derived from NIR spectra \cite{2018ApJ...861..119D,2018A&A...619A.102D}, in $J$, $H$, and $K$ (presented in Supplementary Table~2). 

\begin{figure*}
 \centering
 \includegraphics[width=\textwidth]{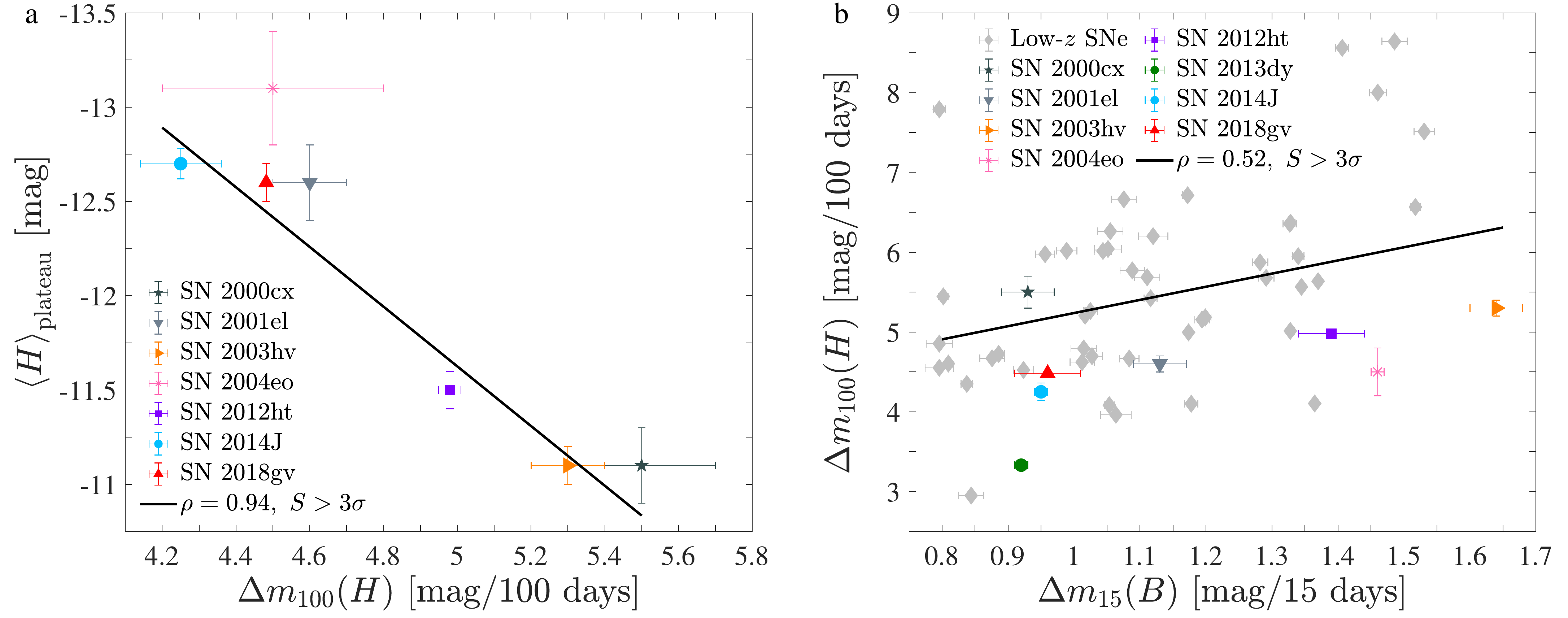}
 \caption{Correlations between NIR light-curve properties. \textbf{a} SNe Ia with steeper decline rates after the second $H$-band maximum, $\Delta m_{100}(H)$, have a fainter average magnitude, $\left<H\right>$, during the plateau phase than SNe with shallower light curves (with a Pearson's correlation coefficient of $\rho=0.94$ and a significance of $S>3\sigma$; solid curve). It is as yet unclear if the SNe fall into two distinct groupings in $\left<H\right>$ or whether there is a continuous distribution. \textbf{b} $\Delta m_{100}(H)$, in turn, is correlated with the intrinsic brightness of the SNe, parameterized by $\Delta m_{15}(B)$. The control sample of low-redshift $(z)$ SNe Ia (gray diamonds) has a Pearson's correlation coefficient of $\rho=0.52$, with a significance of $S>3\sigma$ (solid curve). Error bars represent $1\sigma$ measurement uncertainties. The data behind this figure are provided in Supplementary Data 2 and 3.}
 \label{fig:corr}
\end{figure*}

Our synthesis reveals that the ``IR catastrophe'' does not manifest in the NIR. Moreover, we report the discovery of a plateau in the $H$ band. Hints of this plateau were noted in sparse observations of SNe 1998bu, 2000cx and 2001el but were not explored further \citep{2004A&A...428..555S,2007A&A...470L...1S,2004A&A...426..547S}. Here, we show that this plateau sets in at $\approx 150$ days and lasts for $\approx 250$--$350$ days. While $H$- and \textit{F160W}-band measurements of SNe 2000cx, 2001el, and 2012ht show a flat light curve during discrete portions of this phase range, \textit{F160W} measurements of SNe 2017erp and 2018gv, which span the entire phase range from 150 to 500 days, show that the SNe decline by $\approx 0.3~{\rm mag/100~days}$ during this time.

The $H$-band plateau is also apparent in the $J$ band. Although in this filter we can only compare ground-based data from five SNe from the literature and synthetic photometry of SN 2014J, it seems that the $J$-band plateau is similar to the one observed in the $H$ band: it sets in at roughly $150$ days and also ends at $\approx 400$--$500$ days. 

Most of the literature SNe were not followed long enough in the $K$ band to probe the $150$--$500$ day range of the plateau. However, our synthetic $K$-band photometry of SN 2014J show no evidence of a plateau in this band. Instead, the SN declines at a rate of $1.4 \pm 0.1~{\rm mag/100~days}$, which is consistent with the decline rate in the optical at the same time ($1.299\pm 0.005~{\rm mag/100~days}$, based on data from ref.~\cite{2019ApJ...870...14G}).

In both $J$ and $H$, the plateau spans a range of $\sim 2$ mag, i.e., a factor of 6 in flux. It is instructive that in both filters, the plateau sets in at $\approx 150$ days, even though the SNe had different decline rates in these bands prior to the plateau setting in. At the start of the $J$-band plateau, the SNe have declined by 6--8 mag since peak, while in the $H$ band, they have declined by 5--7 mag. Together with the range of brightnesses the SNe exhibit while on the plateau, this means that the onset of the plateau is probably tied to the age of the SNe and not to other factors, such as their intrinsic brightness.

SNe Ia with higher intrinsic brightnesses and shallower decline rates (parameterized by their $\Delta m_{15}(B)$ values and decline rates in the range $30$--$100$ days, $\Delta m_{100}(H)$) have a higher average brightness, $\left<H\right>$, during the plateau phase than SNe with steeper light curves. Limited by a small number of events in the late-time SN Ia photometry sample, it is unclear whether this correlation divides the sample into two distinct populations or whether there is a continuous distribution in $\left<H\right>$. 

In Figure~3, we show that $\left<H\right>$ is correlated with (and perhaps driven by) $\Delta m_{100}(H)$ at a $>3\sigma$ confidence level (Pearson's $\rho=0.94$, $p=0.002$). A linear fit to the data gives $\left<H\right>=(1.6\pm0.1) \Delta m_{100}(H)-(19.5\pm0.5)~{\rm mag}$. $\Delta m_{100}(H)$, in turn, is correlated with the intrinsic brightness of SNe Ia, often parameterized by $\Delta m_{15}(B)$, the number of magnitudes the SN has declined 15 days after reaching peak in the $B$ band. Using a control sample of normal, low-redshift SNe Ia \cite{2019arXiv190203261A}, we show that the two parameters are correlated at a $>3\sigma$ confidence level ($\rho=0.52$, $p=1.8\times10^{-4}$). The strength of this correlation drops to $>2\sigma$ ($\rho=0.44$, $p=0.0039$) if we remove the five SNe with $\Delta m_{100}(H)>7~{\rm mag/100~days}$ (SNe 2006D, 2006le, 2006lf, 2007st, and 2008hs). Fitting the low-redshift supernova sample, this correlation can be expressed with the linear fit $\Delta m_{100}(H)=(1.649\pm 0.001) \Delta m_{15}(B)+(3.589\pm 0.001)~{\rm mag}$. Based on this correlation, we would have expected SNe 2000cx (2004eo) to have lower (higher) $\left<H\right>$ values than the ones we measure; these peculiar SNe might not be subject to the same correlations as normal SNe Ia \cite{2007MNRAS.377.1531P,2001PASP..113.1178L}.

At $\approx 400$--$500$ days (400 days for the SNe Ia on the upper, brighter $\left<H\right>$, branch and 500 days for the SNe Ia on the lower branch), the SNe leave the plateau and enter a second decline phase. When taken together, the $500<t<800~{\rm days}$ photometry of SNe 2003hv, 2012cg, and 2017erp yields a linear decline rate of $1.15 \pm 0.02~{\rm mag/100~{\rm days}}$, which is broadly consistent with the decline rate of $0.975~{\rm mag/100~{\rm days}}$ expected from the radioactive decay of $^{56}$Co$\to^{56}$Fe, with a half-life of $77.2~{\rm days}$. This is consistent with the optical component of the light curve, which has also been shown to decline in a manner consistent with the radioactive decay of $^{56}$Co during this time. To illustrate this similarity, we plot the fraction of NIR flux out of the combined optical+NIR flux (i.e., the combination of flux from the broad \textit{F350LP} optical filter and the NIR \textit{F160W} filter). Figure~4 shows that, during the plateau, while the SNe continue to fade in the optical, the fraction of light emitted in the NIR steadily increases to $\sim 40$\%. Once the SNe transition off the plateau, this fraction remains constant.

Individually, SNe 2003hv, 2012cg, and 2017erp have different decline rates after the plateau phase: $0.89\pm 0.04$, $1.1 \pm 0.4$, and $2.0 \pm 0.1~{\rm mag/100~days}$, respectively. A larger sample of SNe is required to ascertain whether these differences are due to shot noise or to an intrinsic dispersion of decline rates, such as the dispersion in $\Delta m_{100}(H)$ shown in Figure~3.

After a gap in the NIR measurements between 800 and 1100 days, the only $J$- and $H$-band measurements at $>1100$ days come from SN 2011fe, which was observed in \textit{F125W} at $\sim 1200$--$1500$ days \citep{2017MNRAS.472.2534K} and in \textit{F160W} at $\sim 1100$--$1850$ days \citep{2017ApJ...841...48S}. Unfortunately, SN 2011fe was only observed sparsely along the plateau in the $J$ band, with no similar $H$-band observations. Moreover, it was not observed at all in the $K$ band. Yet the available data are instructive, as they show that the late-time light curve of SN 2011fe slows down in the NIR and proceeds to decline at the same rate ($\approx 0.3~{\rm mag/100~days}$) as in the optical ($\approx 0.3$, $0.4$, and $0.3~{\rm mag/100~days}$ in \textit{F438W}, \textit{F555W}, and \textit{F600LP}, respectively, based on data from ref.~\cite{2017ApJ...841...48S}).

Along with the \textit{HST} imaging, we have also acquired ground-based NIR spectra of SN 2017erp during the plateau phase as well as a \textit{HST} \textit{G141} grism spectrum at 605 days. In Figure~5, we compare the spectra and identify iron-group emission lines that could produce the observed emission features in the spectra. Although we label the features with neutral, singly- and doubly-ionized elements, previous works have shown that toward the end of the plateau, the doubly-ionized lines fade away and the emission features are dominated by singly-ionized lines \cite{2018ApJ...861..119D,2018A&A...619A.102D,2018MNRAS.477.3567M}. We identify emission features at $1.35$--$1.4$ and $1.4$--$1.5~\mu$m that, from among the iron-group elements, can be associated with [Fe I]. We find this identification plausible.

\begin{figure}
 \centering
 \includegraphics[width=0.475\textwidth]{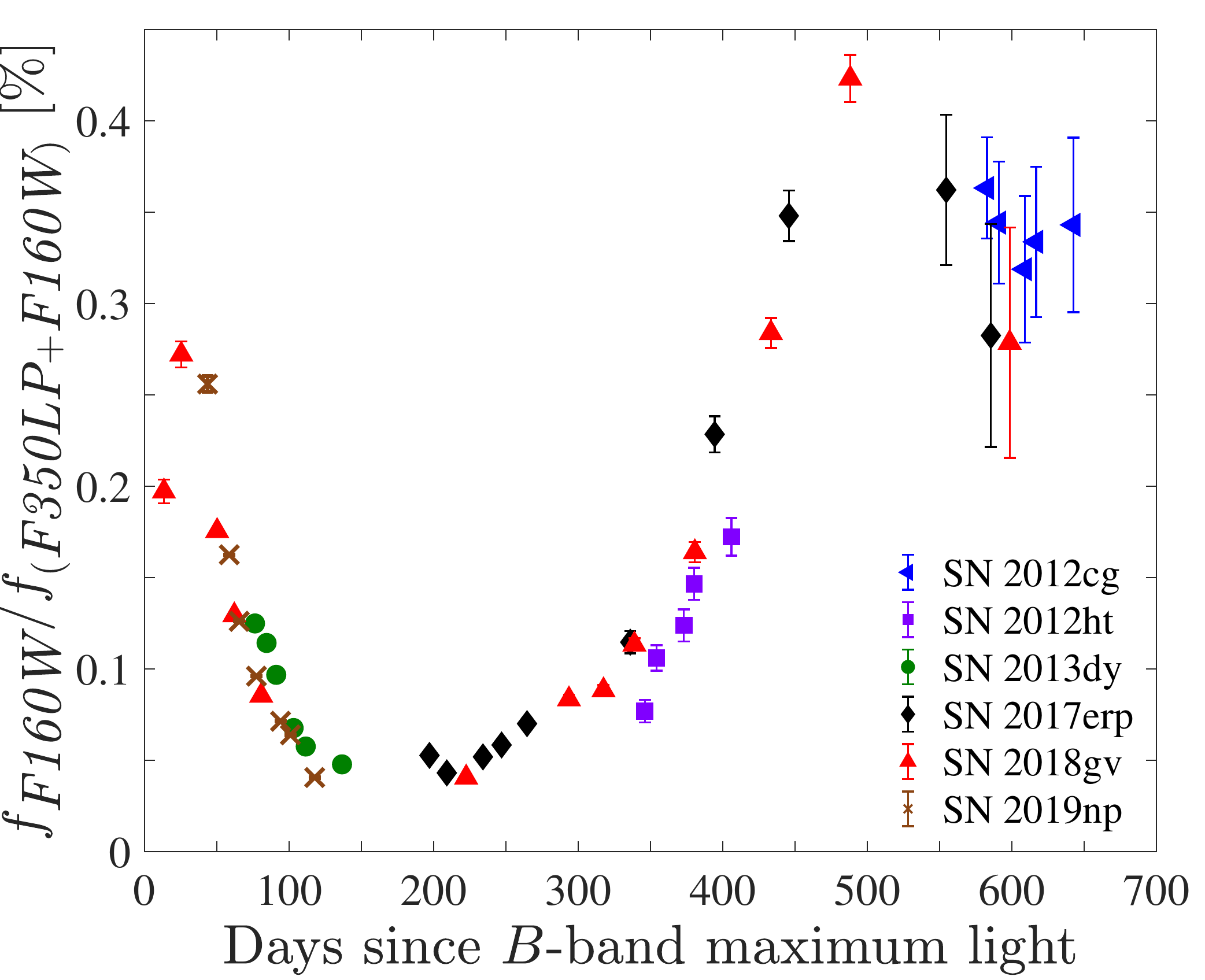}
 \caption{Fraction of NIR light out of the combined optical (\textit{F350LP}) and NIR (\textit{F160W}) flux. During the plateau phase ($150<t<500$ days), the NIR fraction rises to $\sim 40$\% as the optical light curve continues to fade. At $\sim 500$ days, the NIR light curve begins to decline at a rate similar to that of the optical light curve, consistent with the radioactive decay of $^{56}$Co, and the NIR fraction levels off. Error bars represent $1\sigma$ measurement uncertainties. The data behind this figure are provided in Supplementary Data 4.}
\end{figure}

\begin{figure*}
 \centering
 \includegraphics[width=0.97\textwidth]{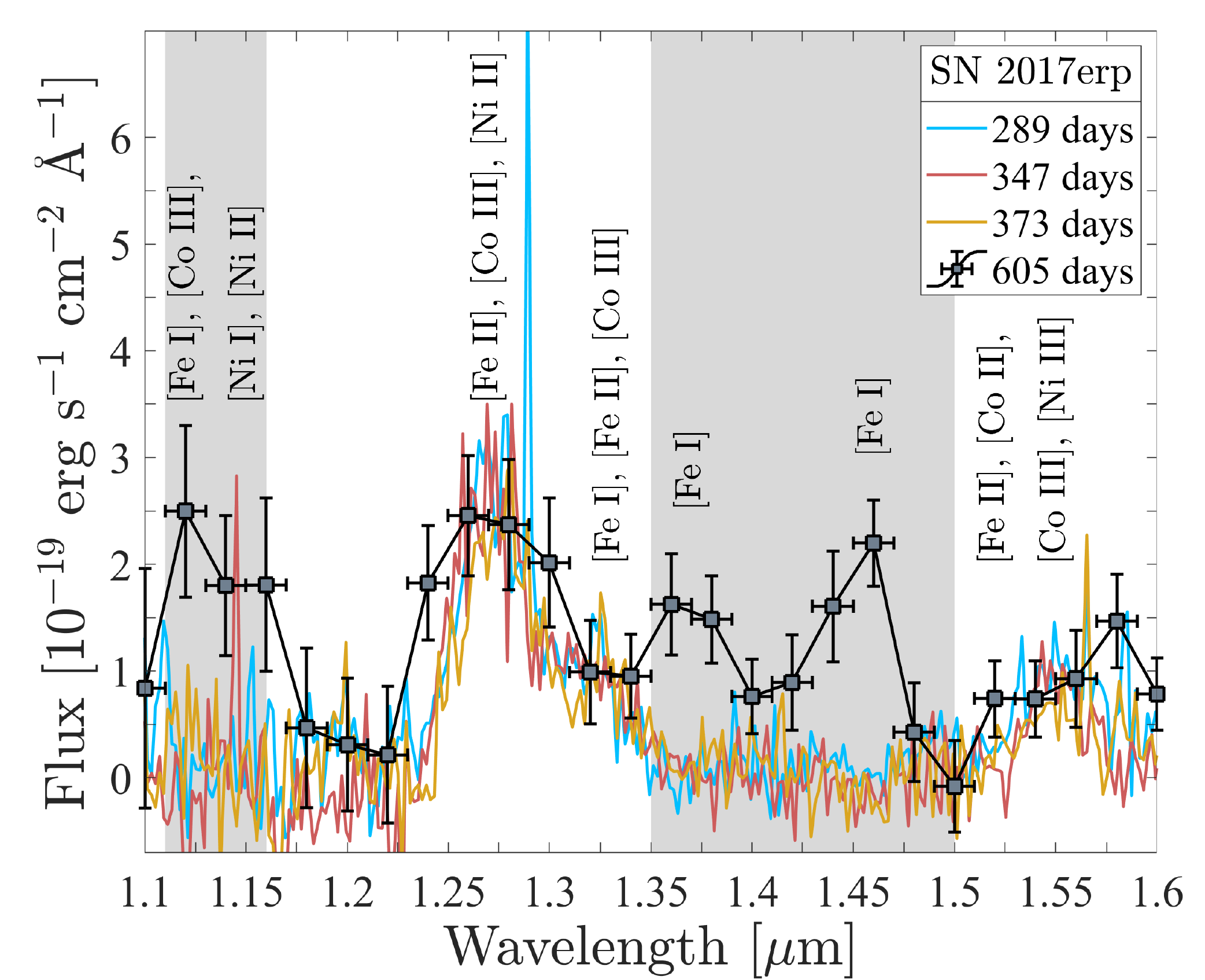}
 \caption{\textit{HST} grism spectrum of SN 2017erp at 605 days (symbols), compared to ground-based spectra of the SN at 289 (blue), 347 (red), and 373 days (yellow). The ground-based spectra have been scaled down by matching the spectra at 347 and 605 days at $1.26~\mu$m. The gray patches denote regions of atmospheric absorption. We identify plausible spectral features at $1.35$--$1.5~\mu$m that could be associated with [Fe I]. The grism spectrum of SN 2017erp is provided in Supplemental Data 5. The ground-based spectra shown here will be published in a future paper. Error bars represent $1\sigma$ measurement uncertainties.}
 \label{fig:spec}
\end{figure*}

\begin{figure*}
 \centering
 \includegraphics[width=\textwidth]{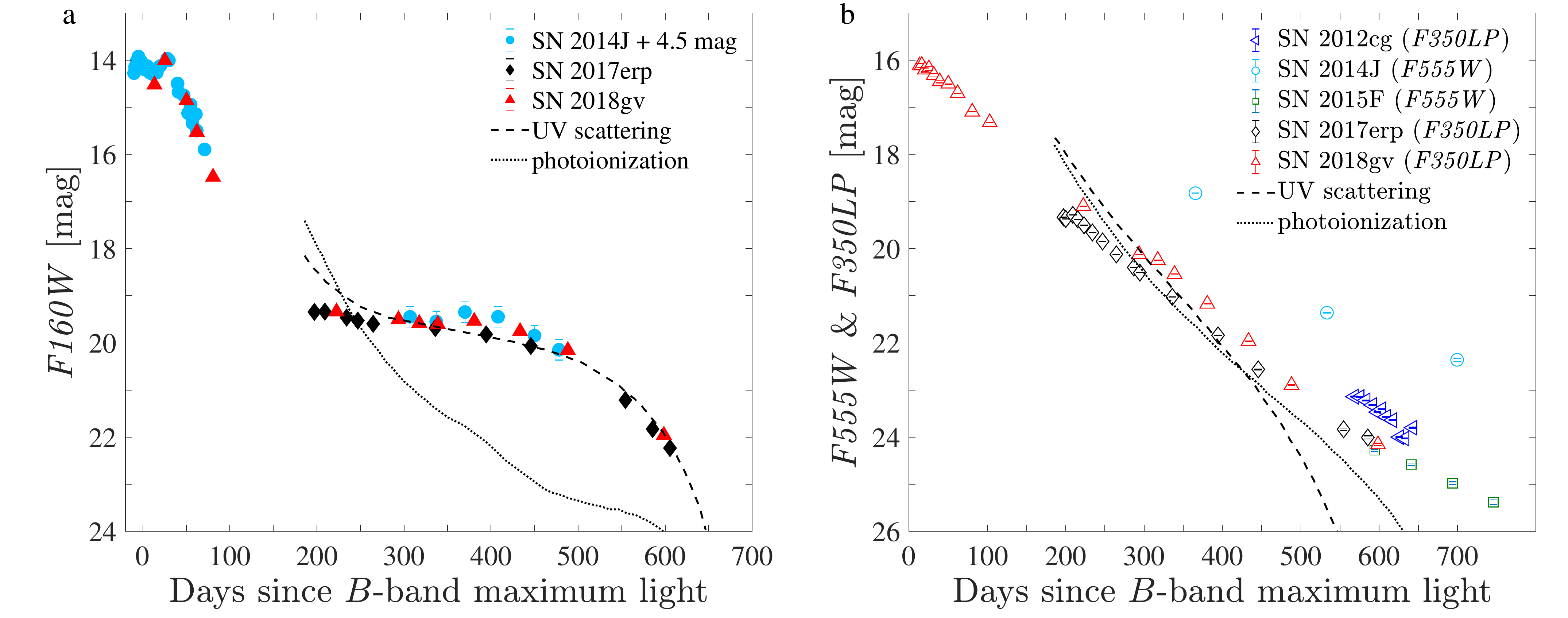}
 \caption{A theoretical explanation for the NIR plateau. \textbf{a} The $H$-band light curves of SNe 2014J, 2017erp, and 2018gv are consistent with an emission model that assumes that photons emitted in the UV are scattered to longer wavelengths (dashed curve), leading to a shift of emission from Fe lines in the optical to [Fe II] lines in the NIR \citep{2004A&A...428..555S}. \textbf{b} Although this model provides a good fit to the NIR data, its assumption that some fraction of the optical emission is lost to the NIR is not borne out by optical light curves, which are consistent with the decline expected from the radioactive decay of $^{56}$Co. The addition of non-local radiative transfer to this model may solve this issue. In both panels, the models have been scaled to fit the \textit{F160W} photometry of SN 2018gv at 300 days. Error bars represent $1\sigma$ measurement uncertainties.}
 \label{fig:Hmodel}
\end{figure*}

The emission features at $1.35$--$1.4$ and $1.4$--$1.5~\mu$m fall within atmospheric absorption bands, which could explain why they were not detected in ground-based spectra at earlier phases. However, the appearance of [Fe I] lines in the NIR should be accompanied by similar lines in the optical (two features at $0.58$ and $0.64~\mu$m in optical spectra of SN 2011fe obtained at 981 and 1034 days have been attributed to [Fe I] but could also be due to other elements, such as hydrogen, oxygen, and calcium; \cite{2015MNRAS.454.1948G,2015MNRAS.448L..48T}) and mid-IR (e.g,, at $24~\mu$m), which would be visible with the \textit{James Webb Space Telescope}.

We leave a detailed modeling of the spectrum to a future paper. If real, the appearance of [Fe I] emission features as strong as the previously dominant [Fe II] feature at $1.26~\mu$m, following the shift in dominance from [Fe III] to [Fe II] along the plateau, would indicate that the transition from the plateau to the second decline phase could be the result of redistribution of the NIR emission to different wavelengths. This could be due to a change in the dominant ionization state of the atoms responsible for the cooling of the ejecta. 

\subsection{Explaining the NIR plateau}

We next review several possible explanations for the occurrence of the NIR plateau.

Light echoes, which are created when light from the SN is scattered by a distant dust sheet toward Earth, have been shown to flatten the light curves of SNe Ia \citep{2005MNRAS.357.1161P,2006MNRAS.369.1949P,2012PASA...29..466R}. For several reasons, we can reject this scenario. First, of the full SN sample shown in Figure~2, only SN 1998bu is known to have a light echo \citep{2001ApJ...549L.215C}, and the effects of that echo have been removed from the photometry presented here \citep{2004A&A...426..547S}. Second, a light echo causes a blueward shift in the light of the SN, which is inconsistent with the reddening of the light curve seen here during the plateau phase. Third, a light echo explanation is inconsistent with the transition of the NIR light curve from the plateau onto the second decline phase at $\approx 500$ days. Finally, a light echo should have impacted the optical portion of the light curve, which in our data remains unperturbed.

We also rule out the effect of dust formed by the SNe. While such dust could conceivably produce an increase in the NIR flux, both dust-formation simulations and observations of SN remnants imply that SNe Ia do not produce substantial ($\gtrsim 10^{-2}~M_\odot$) amounts of dust \citep{2006ApJ...642L.141B,2012MNRAS.420.3557G}, though simulations in which CO and SiO molecules do not form in the ejecta can lead to dust masses as high as $0.2~M_\odot$ \cite{2011ApJ...736...45N}. Observations of SN Ia remnants such as Kepler or Tycho are consistent with the majority of the warm dust being swept up from the circumstellar or interstellar medium \cite{2001A&A...373..281D,2012ApJ...755....3W}. However, at the epochs of our observations, the amount of swept-up dust will be negligible. Furthermore, we would have expected the optical light curves to suffer from dust extinction and change slope during the NIR plateau phase. While such an effect has been observed in some core-collapse SNe \citep{2008MNRAS.389..141M,2009MNRAS.392..894A}, it is not supported by the observations presented here. Finally, the flux density of newly formed dust is expected to be higher in $K$ than in $J$ or $H$ \cite{2011ApJ...736...45N}, contrary to the observations shown in Figure~2.

A potential theoretical explanation for the NIR plateau emerges from several works that have attempted to resolve the conflict between the IR catastrophe, which arises from theoretical modeling of the cooling of the SN ejecta, and the nebular-phase light curves and spectra. In Figure~6, we compare our NIR and optical observations with the only currently available late-time NIR light curve models. Originally compared to the nebular-phase photometry of SN 2000cx \citep{2004A&A...428..555S}, these models were designed to test whether photons emitted in the UV led to high-energy photoionization lines or whether the UV photons were scattered by the many overlapping lines in the UV and shifted to longer wavelengths. These options were tested by turning photoionization on or off in the emission model. In Figure~6, we find that the model that assumes UV scattering is in excellent agreement with our late-time \textit{F160W} photometry, and even predicts the transition from the plateau phase to the second decline phase at $\sim 500$ days. According to this model, the plateau in the NIR is due to optical iron-line emission being shifted into the NIR [Fe II] lines. However, this should then lead to a steeper decline in the optical light curves, which is inconsistent with our observations as well as other optical observations from the literature.

A possible solution to the lack of change in the slope of the optical light curve included the effects of non-local radiative transfer \citep{2015ApJ...814L...2F}. The updated model, which was compared to an optical spectrum of SN 2011fe taken at 1034 days after the explosion \citep{2015MNRAS.448L..48T}, found that without non-local radiative transfer, only $\sim 5\%$ of the energy deposited into the ejecta emerged in the optical and NIR, while $\sim 15\%$ emerged in the UV and the rest in the mid-IR. The addition of non-local radiative transfer converted most of the UV light into the optical and NIR, thus correcting for the decline in the optical light curve predicted by the IR catastrophe.

Unfortunately, ref.~\cite{2015ApJ...814L...2F} did not show the effects of the updated emission model on the NIR light curve. Thus, it remains to be seen whether an emission model that includes both UV scattering and non-local radiative transfer can reproduce both the NIR plateau and the concomitantly constantly-declining optical light curve. 

The explanation for the occurrence of the NIR plateau offered by UV scattering is reminiscent of the explanation put forth for the occurrence of the second NIR peak at $\approx 20$--$30$ days \cite{2006ApJ...649..939K}. As the ejecta cool, a ``recombination wave'' from doubly- to singly-ionized iron-group elements seemingly propagates inwards through the ejecta towards the iron-rich core. This ``wave'' is the result of two competing factors: the outer part of the ejecta expands more than the inner part and so is cooler at any given time, but the inner part is denser; thus, recombination is faster and occurs at higher temperatures. The low ionization state, iron-rich gas in the core has a very high opacity to UV photons due to line-blocking from atomic resonance transitions, which leads to a redistribution of radiation from the UV to the NIR via fluorescence \cite{2000ApJ...530..757P,1995ApJ...443...89H}. This explanation is consistent with the observed transition from doubly- to singly-ionized forbidden lines along the plateau as well as the appearance of neutral [Fe I] lines---signaling a shift from $2\to1$ recombination to $1\to0$---once the plateau ends. Thus, the plateau phase at 150--500 days can be thought of as an additional, drawn-out peak in the NIR light curve.

The fluorescence mechanism considered above also produces correlations between the brightness and timing of the second NIR peak and the mass of $^{56}$Ni created during the explosion (which is responsible for the overall luminosity of the SN, often parameterized by $\Delta m_{15}(B)$ \cite{1995ApJ...455L.147N}). Applying the fluorescence explanation to the NIR plateau thus naturally explains the correlations shown in Figure~3: SNe Ia that produce larger amounts of $^{56}$Ni are hotter and have larger iron cores, which together lead to a later development of the $2\to1$ recombination wave along with a larger, more luminous fluorescent shell. This creates a delayed, brighter second NIR peak. Similarly, these SNe Ia would then be expected to have brighter plateaus, as seen in our data. It remains to be seen whether a larger sample of SNe observed along the plateau would also reveal a difference in the timing of its onset and demise, as might be expected from this model.


\section{Discussion}
\label{sec:discuss}

We have presented late-time \textit{HST} imaging of five SNe Ia out to 605 days past $B$-band maximum light. We report a year-long plateau in the $J$- and $H$-band light curves that spans $\approx 150$--$500$ days. This plateau, together with the concomitantly constant decline rate of the optical light curve, are consistent with a masking of the infrared catastrophe in the optical and NIR by scattering of UV photons to longer wavelengths. Further observations are required to establish the correlation shown here between the intrinsic luminosity of the SNe and their average $H$-band magnitudes on the plateau, and to fill in the 150--500 day period in the $K$ band, where synthetic photometry of SN 2014J implies a decline rate similar to the optical. We also present a NIR spectrum of SN 2017erp at $>500$ days, in which we report tentative identifications of [Fe I] emission features. A detailed modeling of the spectrum and a larger sample of NIR spectra taken at $>500$ days, together with optical and mid-IR spectra, is necessary to validate these features, which hint that the end of the plateau phase coincides with a shift in the dominant ionization state of the SN ejecta.

Our discovery and characterization of the NIR plateau and second decline phase shed new light on the complex radiative transfer processes that take place in SNe Ia. The data presented here will also assist in the use of SNe Ia as standard candles. Recent work has shown that SNe Ia are most standard in the NIR, and especially in the $H$ band. Our results indicate that NIR cosmology surveys (e.g., ref.~\cite{2019arXiv190203261A}) may best be served by obtaining template images of their SNe after they have fallen off the plateau and by taking into account the $\approx2$ mag spread in brightness along the plateau and its correlation with the intrinsic brightness of the SNe.


\section{Methods}
\label{sec:methods}

\subsection{Imaging}
\label{subsec:imaging}

We imaged SNe 2012ht, 2013dy, 2017erp, 2018gv, and 2019np with \textit{HST}/WFC3 in the \textit{F350LP}, \textit{F555W}, \textit{F814W}, and \textit{F160W} filters through programs GO--12880, 15145, 15640, 15686, and 15693. We used the \texttt{Tweakreg} and \texttt{AstroDrizzle} tasks included in the \href{http://drizzlepac.stsci.edu/}{\texttt{DrizzlePac}} Python package \citep{2012AAS...22013515H} to align the \textit{HST} images and remove cosmic rays and bad pixels. 

Next, we performed point-spread-function (PSF) fitting photometry using \href{http://americano.dolphinsim.com/dolphot/}{\texttt{Dolphot}} \citep{2000PASP..112.1383D}. For the optical images, we used the \texttt{flc} files produced by the \textit{HST} WFC3 pipeline, which are corrected for charge transfer efficiency effects; for the NIR \textit{F160W} images, we used the \texttt{flt} files, since the IR channel of WFC3 does not suffer from charge transfer effects. 

Because GO--12880, 15145, and 15640 were designed not to observe the SNe but to discover variable stars in the host galaxies, \texttt{Dolphot} reported the SNe as saturated in many of the optical images. Though some of the pixels at the location of the SN are saturated, most of the PSF is intact and can be sampled with aperture photometry. The resultant photometry will be fainter than the true values, but not by a wide margin. Thus, in order to estimate the brightness of the SNe in the saturated optical visits, we also performed aperture photometry with a $0.4^{\prime\prime}$ aperture. For the latter, we applied the latest \href{http://www.stsci.edu/hst/wfc3/phot_zp_lbn}{aperture corrections} \cite{2016wfc..rept....3D}. On average, the PSF-fitting photometry is systematically brighter than the aperture photometry by $\sim0.1$ mag. Both sets of photometry are presented in Supplementary Table~1. Crucially, saturation was not an issue in the NIR, since WFC3/IR conducts many sub exposures that are then combined to form the final images. We stress that the optical photometry reported here was not used for the majority of our analysis and that results in saturated images should be treated as lower limits.

\subsection{Spectroscopy}
\label{subsec:spec}

Three spectra of SN 2017erp were obtained with VLT+XShooter at 292, 349, and 375 days. XShooter is an echelle spectrograph with three arms (UVB, VIS, NIR) covering the wavelength range $\sim3000$--$25000~$\AA\ \citep{2011A&A...536A.105V}. We used a setup with slit widths of $0.8^{\prime\prime}$, $0.9^{\prime\prime}$, and $0.9^{\prime\prime}$ in the UVB, VIS, and NIR arms, respectively. The spectra were reduced using the \texttt{REFLEX} pipeline to produce flux-calibrated one-dimensional spectra \citep{2018MNRAS.477.3567M,2010SPIE.7737E..28M,2013A&A...559A..96F,2016MNRAS.457.3254M}. Telluric corrections have been applied to the optical spectra using the \texttt{MOLECFIT} package \citep{2015A&A...576A..78K,2015A&A...576A..77S}. The spectra were absolute flux-calibrated using optical photometry obtained by the Las Cumbres Observatory Global Telescope array at similar epochs \cite{2013PASP..125.1031B,2016MNRAS.459.3939V}. 

Through \textit{HST} program GO--15686, we obtained a WFC3 \textit{G141} grism spectrum of SN 2017erp on 2019 Feb. 26, at a phase of 605.5 days. We extracted the one-dimensional spectrum from $1.0$--$1.7~\mu$m in bins of 200~\AA\ using sparse-matrix techniques \cite{2018PASP..130c4501R}, but since only a single orient was available, we required additional steps to remove the contamination from the host galaxy \cite{2017Natur.551...71T}. Next, we matched the emission features of the spectrum with iron-group forbidden lines from the National Institute of Standards and Technology's \href{https://physics.nist.gov/PhysRefData/ASD/lines_form.html}{Atomic Spectra Database} and the compilation of atomic data from \texttt{CMFGEN} \citep{1998ApJ...496..407H,2012MNRAS.424..252H,2016MNRAS.462.3350T}, including forbidden line data for [Fe II] \cite{1996A&AS..120..361Q} and [Fe III] \cite{1996A&AS..116..573Q}.

Late-time NIR photometry of SN 2014J was acquired by synthetically photometering NIR spectra taken at 307--478 days \cite{2018ApJ...861..119D,2018A&A...619A.102D}. The spectra from ref.~\cite{2018A&A...619A.102D} were shared with us by Suhail Dhawan, while the spectra from ref.~\cite{2018ApJ...861..119D} were re-reduced by us. We retrieved the raw Gemini Near-InfraRed Spectrograph data from the Gemini archive and applied a standard reduction using the Gemini package in \texttt{Iraf}. The observations were conducted in cross-dispersed mode, and the reduction consisted of flat-fielding, sky-subtraction using observations at alternate dither positions, wavelength calibration using an arc lamp, geometric correction of each spectroscopic order, and extraction to a one-dimensional spectrum. A telluric standard, HIP 32549, had been observed immediately prior to each spectrum. At each epoch, we derived the telluric corrections and detector sensitivity function by normalising the reduced spectra of HIP 32549 to a 9700 K blackbody curve (appropriate for the spectral type A0V), scaled to the catalogued $JHK$ magnitudes from 2MASS \cite{2006AJ....131.1163S}. Flux calibration was achieved by dividing the SN spectrum by this function. We caution that this approach does not account for differential slit losses between the observations of the SN and the standard, adding some uncertainty to the absolute flux scale. However, as these observations were obtained close together in both time and airmass it is likely that any differences in atmospheric seeing (and hence associated slit losses) were minimal. To account for this, we attribute an uncertainty of 20\% to the synthetic photometry.

\subsection{Synthesis with ground-based observations}
\label{subsec:synthesis}

In Figure~2, we synthesize our late-time PSF-fitting \textit{F160W} photometry with early-phase ground-based $H$-band photometry of SNe 2012ht \citep{2018ApJ...869...56B} and 2013dy \citep{2015MNRAS.452.4307P}. We then compare the resultant light curves to SNe 1998bu \citep{2004A&A...426..547S,1998IAUC.6907....2M,1999ApJS..125...73J,2000MNRAS.319..223H,2000MNRAS.314..782M}, 2000cx \citep{2003PASP..115..277C,2004A&A...428..555S}, 2001el \citep{2003AJ....125..166K,2007A&A...470L...1S}, 2003hv \citep{2009A&A...505..265L}, 2004eo \citep{2007MNRAS.377.1531P}, 2011fe \citep{2017ApJ...841...48S,2012ApJ...754...19M}, 2012cg \citep{2016ApJ...819...31G,2016ApJ...820...92M}, and 2014J \citep{2016ApJ...822L..16S}.

To overcome any systematics associated with measurements of the distance moduli and host-galaxy extinctions of the SNe by the various groups cited above, we do not correct the photometry of the SNe for any of these effects. Instead, motivated by the finding that SNe Ia have a low dispersion of $\sim 0.1$ mag around peak in \textit{F160W} \citep{2019arXiv190203261A}, we use the early-phase portions of their light curves to scale them to SN 2011fe. Applying a distance modulus to SN 2011fe of $28.86$ mag \citep{2012ApJ...754...19M}, the combined light curves peak at $\approx -18.3$ mag, consistent with the template \textit{F160W} peak magnitude derived by ref.~\cite{2019arXiv190203261A}. The same scaling factors are also applied to the $J$- and $K$-band data presented in Figure~2. For SNe 2017erp and 2019np, which do not have early-time NIR data, we use distance moduli of 32.29 and 32.58 mag, respectively \citep{2019ApJ...877..152B,2016AJ....152...50T} and Galactic reddenings in \textit{F160W} of $0.055$ and $0.010$ mag, respectively \citep{2011ApJ...737..103S}. SN 2018gv, which does not have coverage of the first $H$-band peak, was matched to SN 2011fe at 25 days  \cite{2019arXiv190310820Y} using early-time optical photometry.

In this work, we do not apply $S$-corrections between the ground-based $H$-band photometry and space-based \textit{F160W} photometry. These are difficult to measure, since most of the SNe we analyze here do not have synchronous $H$ and \textit{F160W} data. To estimate this effect, we performed synthetic aperture photometry on a 425-day spectrum of SN 2013aa \citep{2018MNRAS.477.3567M} and found that the resultant \textit{F160W} photometry was $\sim 0.2$ mag fainter than in the $H$ band. Though larger than the formal uncertainties of the photometry in Supplementary Table~1, this effect is an order of magnitude smaller than the dispersion of magnitudes we measure during the plateau phase.  

\subsection{Comparison with theory}
\label{subsec:lcfits}

We extracted the SN Ia emission models shown in Figure~6 by digitizing the curves from figure 11 of ref.~\cite{2004A&A...428..555S} with \href{https://automeris.io/WebPlotDigitizer}{\texttt{WebPlotDigitizer}}. Next, we scaled the $H$- and $V$-band models to the \textit{F160W} and \textit{F350LP} photometry of SN 2018gv at 300 days, respectively. For display purposes, we also scaled the synthetic $H$-band photometry of SN 2014J by $4.5$ mag. We emphasize that the theoretical models were only scaled---not fitted---to the photometry of SN 2018gv.


\section*{Acknowledgments}

We thank Gabriel Brammer, Daniel Eisenstein, Wolfgang Kerzendorf, Stuart Sim, Lou Strolger, and the referees for helpful discussions and comments; Suhail Dhawan for sharing his spectra of SN 2014J; and Annalisa Calamida, Weston Eck, Mario Gennaro, and William Januszewski for supporting the \textit{HST} programs used here. 

\textbf{Funding:} O.G. is supported by an NSF Astronomy and Astrophysics Fellowship under award AST-1602595. K.M. acknowledges support from H2020 through an ERC Starting Grant (758638). M.N. is supported by a Royal Astronomical Society Research Fellowship. R.T.F. acknowledges support from NASA ATP award 80NSSC18K1013. This work is based on data obtained with the NASA/ESA {\it Hubble Space Telescope}, all of which was obtained from MAST. Support for MAST for non-\textit{HST} data is provided by the NASA Office of Space Science via grant NNX09AF08G and by other grants and contracts. Based on data taken at the European Organization for Astronomical Research in the Southern Hemisphere, Chile, under program IDs: 0100.D-0242(A) and 0101.D-0443(A). This research has made use of NASA's Astrophysics Data System and the NASA/IPAC Extragalactic Database (NED) which is operated by the Jet Propulsion Laboratory, California Institute of Technology, under contract with NASA. The NIST databases were funded, in part, by NIST's Standard Reference Data Program (SRDP) and by NIST's Systems Integration for Manufacturing Applications (SIMA) Program. Finally, this work has made use of the Open Supernova Catalog \citep{2017ApJ...835...64G}. 

\textbf{Author contributions:} O.G. planned the observations for \textit{HST} programs GO--15686 and 15693, reduced the \textit{HST} imaging data, performed the analysis, and wrote the manuscript. K.M. obtained the ground-based spectra of SN 2017erp. M.N. reduced the Gemini spectra of SN 2014J. R.R. reduced the grism observation of SN 2017erp obtained through \textit{HST} program GO--15686. A.A. measured $\Delta m_{100}(H)$ values for the correlation study. A.G.R. planned the \textit{HST} observations for programs GO--12880, 15145, and 15640. I.R.S., R.F., and L.S. assisted with the theoretical analysis of the observations. 

\textbf{Competing interests:} The authors declare that there are no competing interests. 

\textbf{Data availability:} \textit{HST} observations are available through the Mikulski Archive for Space Telescopes (MAST). The data for Supplementary Tables~1--2 and Figures~2--5 are provided in the Supplemental Material section as machine-readable tables. Any further data are available from the corresponding author upon reasonable request.

\section*{Supplementary Material}

\noindent Supplementary Table 1: \textit{Hubble Space Telescope} photometry of SNe 2012ht, 2013dy, 2017erp, 2018gv and 2019np, presented as a machine-readable table.
\\ \\
Supplementary Table 2: Synthetic photometry of SN 2014J, presented as a machine-readable table.
\\ \\
Supplementary Data 1: Data behind Figure 2, presented as a machine-readable table.
\\ \\
Supplementary Data 2: Data behind Figure 3A, presented as a machine-readable table.
\\ \\
Supplementary Data 3: Data behind Figure 3B, presented as a machine-readable table.
\\ \\
Supplementary Data 4: Data behind Figure 4, presented as a machine-readable table.
\\ \\
Supplementary Data 5: Data behind figure 5: \textit{Hubble Space Telescope} grism spectrum of SN 2017erp, presented as a machine-readable table.




\begin{thebibliography}{}
\expandafter\ifx\csname natexlab\endcsname\relax\def\natexlab#1{#1}\fi

\bibitem[{{Anupama} {et~al.}(2009){Anupama}, {Sahu}, {Gurugubelli}, {Prabhu},
  {Tominaga}, {Tanaka}, \& {Nomoto}}]{2009MNRAS.392..894A}
{Anupama}, G.~C., {Sahu}, D.~K., {Gurugubelli}, U.~K., {et~al.} 2009, \mnras,
  392, 894

\bibitem[{{Arnett}(1982)}]{1982ApJ...253..785A}
{Arnett}, W.~D. 1982, \apj, 253, 785

\bibitem[{{Avelino} {et~al.}(2019){Avelino}, {Friedman}, {Mandel}, {Jones},
  {Challis}, \& {Kirshner}}]{2019arXiv190203261A}
{Avelino}, A., {Friedman}, A.~S., {Mandel}, K.~S., {et~al.} 2019, arXiv
  e-prints, arXiv:1902.03261

\bibitem[{{Axelrod}(1980)}]{1980PhDT.........1A}
{Axelrod}, T.~S. 1980, PhD thesis, California Univ., Santa Cruz.

\bibitem[{{Borkowski} {et~al.}(2006){Borkowski}, {Williams}, {Reynolds},
  {Blair}, {Ghavamian}, {Sankrit}, {Hendrick}, {Long}, {Raymond}, {Smith},
  {Points}, \& {Winkler}}]{2006ApJ...642L.141B}
{Borkowski}, K.~J., {Williams}, B.~J., {Reynolds}, S.~P., {et~al.} 2006, \apjl,
  642, L141

\bibitem[{{Brown} {et~al.}(2019){Brown}, {Hosseinzadeh}, {Jha}, {Sand},
  {Vieira}, {Wang}, {Dai}, {Dettman}, {Mould}, {Uddin}, {Wang}, {Arcavi},
  {Bento}, {Burns}, {Diamond}, {Hiramatsu}, {Howell}, {Hsiao}, {Marion},
  {McCully}, {Milne}, {Mirzaqulov}, {Ruiter}, {Valenti}, \&
  {Xiang}}]{2019ApJ...877..152B}
{Brown}, P.~J., {Hosseinzadeh}, G., {Jha}, S.~W., {et~al.} 2019, \apj, 877, 152

\bibitem[{{Brown} {et~al.}(2013){Brown}, {Baliber}, {Bianco}, {Bowman},
  {Burleson}, {Conway}, {Crellin}, {Depagne}, {De Vera}, {Dilday}, {Dragomir},
  {Dubberley}, {Eastman}, {Elphick}, {Falarski}, {Foale}, {Ford}, {Fulton},
  {Garza}, {Gomez}, {Graham}, {Greene}, {Haldeman}, {Hawkins}, {Haworth},
  {Haynes}, {Hidas}, {Hjelstrom}, {Howell}, {Hygelund}, {Lister}, {Lobdill},
  {Martinez}, {Mullins}, {Norbury}, {Parrent}, {Paulson}, {Petry}, {Pickles},
  {Posner}, {Rosing}, {Ross}, {Sand}, {Saunders}, {Shobbrook}, {Shporer},
  {Street}, {Thomas}, {Tsapras}, {Tufts}, {Valenti}, {Vander Horst}, {Walker},
  {White}, \& {Willis}}]{2013PASP..125.1031B}
{Brown}, T.~M., {Baliber}, N., {Bianco}, F.~B., {et~al.} 2013, \pasp, 125, 1031

\bibitem[{{Burns} {et~al.}(2018){Burns}, {Parent}, {Phillips}, {Stritzinger},
  {Krisciunas}, {Suntzeff}, {Hsiao}, {Contreras}, {Anais}, {Boldt}, {Busta},
  {Campillay}, {Castell{\'o}n}, {Folatelli}, {Freedman}, {Gonz{\'a}lez},
  {Hamuy}, {Heoflich}, {Krzeminski}, {Madore}, {Morrell}, {Persson}, {Roth},
  {Salgado}, {Ser{\'o}n}, \& {Torres}}]{2018ApJ...869...56B}
{Burns}, C.~R., {Parent}, E., {Phillips}, M.~M., {et~al.} 2018, \apj, 869, 56

\bibitem[{{Candia} {et~al.}(2003){Candia}, {Krisciunas}, {Suntzeff},
  {Gonz{\'a}lez}, {Espinoza}, {Leiton}, {Rest}, {Smith}, {Cuadra}, {Tavenner},
  {Logan}, {Snider}, {Thomas}, {West}, {Gonz{\'a}lez}, {Gonz{\'a}lez},
  {Phillips}, {Hastings}, \& {McMillan}}]{2003PASP..115..277C}
{Candia}, P., {Krisciunas}, K., {Suntzeff}, N.~B., {et~al.} 2003, \pasp, 115,
  277

\bibitem[{{Cappellaro} {et~al.}(1997){Cappellaro}, {Mazzali}, {Benetti},
  {Danziger}, {Turatto}, {della Valle}, \& {Patat}}]{1997A&A...328..203C}
{Cappellaro}, E., {Mazzali}, P.~A., {Benetti}, S., {et~al.} 1997, \aap, 328,
  203

\bibitem[{{Cappellaro} {et~al.}(2001){Cappellaro}, {Patat}, {Mazzali},
  {Benetti}, {Danziger}, {Pastorello}, {Rizzi}, {Salvo}, \&
  {Turatto}}]{2001ApJ...549L.215C}
{Cappellaro}, E., {Patat}, F., {Mazzali}, P.~A., {et~al.} 2001, \apjl, 549,
  L215

\bibitem[{{Chevalier}(1976)}]{1976ApJ...207..872C}
{Chevalier}, R.~A. 1976, \apj, 207, 872

\bibitem[{{Colgate} \& {McKee}(1969)}]{Colgate1969}
{Colgate}, S.~A., \& {McKee}, C. 1969, \apj, 157, 623

\bibitem[{{Deustua} {et~al.}(2016){Deustua}, {Mack}, {Bowers}, {Baggett},
  {Bajaj}, {Dahlen}, {Durbin}, {Gosmeyer}, {Gunning}, {Hammer}, {Hartig},
  {Khandrika}, {MacKenty}, {Ryan}, {Sabbi}, \& {Sosey}}]{2016wfc..rept....3D}
{Deustua}, S.~E., {Mack}, J., {Bowers}, A.~S., {et~al.} 2016, {UVIS 2.0
  Chip-dependent Inverse Sensitivity Values}, Tech. rep.

\bibitem[{{Dhawan} {et~al.}(2018){Dhawan}, {Fl{\"o}rs}, {Leibundgut},
  {Maguire}, {Kerzendorf}, {Taubenberger}, {Van Kerkwijk}, \&
  {Spyromilio}}]{2018A&A...619A.102D}
{Dhawan}, S., {Fl{\"o}rs}, A., {Leibundgut}, B., {et~al.} 2018, \aap, 619, A102

\bibitem[{{Diamond} {et~al.}(2018){Diamond}, {Hoeflich}, {Hsiao}, {Sand},
  {Sonneborn}, {Phillips}, {Hristov}, {Collins}, {Ashall}, {Marion},
  {Stritzinger}, {Morrell}, {Gerardy}, \& {Penney}}]{2018ApJ...861..119D}
{Diamond}, T.~R., {Hoeflich}, P., {Hsiao}, E.~Y., {et~al.} 2018, \apj, 861, 119

\bibitem[{{Dimitriadis} {et~al.}(2017){Dimitriadis}, {Sullivan}, {Kerzendorf},
  {Ruiter}, {Seitenzahl}, {Taubenberger}, {Doran}, {Gal-Yam}, {Laher},
  {Maguire}, {Nugent}, {Ofek}, \& {Surace}}]{2017MNRAS.468.3798D}
{Dimitriadis}, G., {Sullivan}, M., {Kerzendorf}, W., {et~al.} 2017, \mnras,
  468, 3798

\bibitem[{{Dolphin}(2000)}]{2000PASP..112.1383D}
{Dolphin}, A.~E. 2000, \pasp, 112, 1383

\bibitem[{{Douvion} {et~al.}(2001){Douvion}, {Lagage}, {Cesarsky}, \&
  {Dwek}}]{2001A&A...373..281D}
{Douvion}, T., {Lagage}, P.~O., {Cesarsky}, C.~J., \& {Dwek}, E. 2001, \aap,
  373, 281

\bibitem[{{Fransson} \& {Jerkstrand}(2015)}]{2015ApJ...814L...2F}
{Fransson}, C., \& {Jerkstrand}, A. 2015, \apjl, 814, L2

\bibitem[{{Freudling} {et~al.}(2013){Freudling}, {Romaniello}, {Bramich},
  {Ballester}, {Forchi}, {Garc{\'{\i}}a-Dabl{\'o}}, {Moehler}, \&
  {Neeser}}]{2013A&A...559A..96F}
{Freudling}, W., {Romaniello}, M., {Bramich}, D.~M., {et~al.} 2013, \aap, 559,
  A96

\bibitem[{{Gomez} {et~al.}(2012){Gomez}, {Clark}, {Nozawa}, {Krause}, {Gomez},
  {Matsuura}, {Barlow}, {Besel}, {Dunne}, {Gear}, {Hargrave}, {Henning},
  {Ivison}, {Sibthorpe}, {Swinyard}, \& {Wesson}}]{2012MNRAS.420.3557G}
{Gomez}, H.~L., {Clark}, C.~J.~R., {Nozawa}, T., {et~al.} 2012, \mnras, 420,
  3557

\bibitem[{{Graham} {et~al.}(2015){Graham}, {Nugent}, {Sullivan}, {Filippenko},
  {Cenko}, {Silverman}, {Clubb}, \& {Zheng}}]{2015MNRAS.454.1948G}
{Graham}, M.~L., {Nugent}, P.~E., {Sullivan}, M., {et~al.} 2015, \mnras, 454,
  1948

\bibitem[{{Graur}(2019)}]{2019ApJ...870...14G}
{Graur}, O. 2019, \apj, 870, 14

\bibitem[{{Graur} {et~al.}(2016){Graur}, {Zurek}, {Shara}, {Riess},
  {Seitenzahl}, \& {Rest}}]{2016ApJ...819...31G}
{Graur}, O., {Zurek}, D., {Shara}, M.~M., {et~al.} 2016, \apj, 819, 31

\bibitem[{{Graur} {et~al.}(2018{\natexlab{a}}){Graur}, {Zurek}, {Cara}, {Rest},
  {Seitenzahl}, {Shappee}, {Shara}, \& {Riess}}]{2018ApJ...866...10G}
{Graur}, O., {Zurek}, D.~R., {Cara}, M., {et~al.} 2018{\natexlab{a}}, \apj,
  866, 10

\bibitem[{{Graur} {et~al.}(2018{\natexlab{b}}){Graur}, {Zurek}, {Rest},
  {Seitenzahl}, {Shappee}, {Fisher}, {Guillochon}, {Shara}, \&
  {Riess}}]{2018ApJ...859...79G}
{Graur}, O., {Zurek}, D.~R., {Rest}, A., {et~al.} 2018{\natexlab{b}}, \apj,
  859, 79

\bibitem[{{Guillochon} {et~al.}(2017){Guillochon}, {Parrent}, {Kelley}, \&
  {Margutti}}]{2017ApJ...835...64G}
{Guillochon}, J., {Parrent}, J., {Kelley}, L.~Z., \& {Margutti}, R. 2017, \apj,
  835, 64

\bibitem[{{Hack} {et~al.}(2012){Hack}, {Dencheva}, {Fruchter}, {Armstrong},
  {Avila}, {Baggett}, {Bray}, {Droettboom}, {Dulude}, {Gonzaga}, {Grogin},
  {Kozhurina-Platais}, {Lucas}, {Mack}, {MacKenty}, {Petro}, {Pirzkal},
  {Rajan}, {Smith}, {Sontag}, \& {Ubeda}}]{2012AAS...22013515H}
{Hack}, W.~J., {Dencheva}, N., {Fruchter}, A.~S., {et~al.} 2012, in American
  Astronomical Society Meeting Abstracts, Vol. 220, American Astronomical
  Society Meeting Abstracts \#220, 135.15

\bibitem[{{Hernandez} {et~al.}(2000){Hernandez}, {Meikle}, {Aparicio}, {Benn},
  {Burleigh}, {Chrysostomou}, {Fernandes}, {Geballe}, {Hammersley},
  {Iglesias-Paramo}, {James}, {James}, {Kemp}, {Lister}, {Martinez-Delgado},
  {Oscoz}, {Pollacco}, {Rozas}, {Smartt}, {Sorensen}, {Swaters}, {Telting},
  {Vacca}, {Walton}, \& {Zapatero-Osorio}}]{2000MNRAS.319..223H}
{Hernandez}, M., {Meikle}, W.~P.~S., {Aparicio}, A., {et~al.} 2000, \mnras,
  319, 223

\bibitem[{{Hillier} \& {Dessart}(2012)}]{2012MNRAS.424..252H}
{Hillier}, D.~J., \& {Dessart}, L. 2012, \mnras, 424, 252

\bibitem[{{Hillier} \& {Miller}(1998)}]{1998ApJ...496..407H}
{Hillier}, D.~J., \& {Miller}, D.~L. 1998, \apj, 496, 407

\bibitem[{{H{\"o}flich}(1995)}]{1995ApJ...443...89H}
{H{\"o}flich}, P. 1995, \apj, 443, 89

\bibitem[{{Jha} {et~al.}(1999){Jha}, {Garnavich}, {Kirshner}, {Challis},
  {Soderberg}, {Macri}, {Huchra}, {Barmby}, {Barton}, {Berlind}, {Brown},
  {Caldwell}, {Calkins}, {Kannappan}, {Koranyi}, {Pahre}, {Rines}, {Stanek},
  {Stefanik}, {Szentgyorgyi}, {V{\"a}is{\"a}nen}, {Wang}, {Zajac}, {Riess},
  {Filippenko}, {Li}, {Modjaz}, {Treffers}, {Hergenrother}, {Grebel},
  {Seitzer}, {Jacoby}, {Benson}, {Rizvi}, {Marschall}, {Goldader}, {Beasley},
  {Vacca}, {Leibundgut}, {Spyromilio}, {Schmidt}, \&
  {Wood}}]{1999ApJS..125...73J}
{Jha}, S., {Garnavich}, P.~M., {Kirshner}, R.~P., {et~al.} 1999, \apjs, 125, 73

\bibitem[{{Kasen}(2006)}]{2006ApJ...649..939K}
{Kasen}, D. 2006, \apj, 649, 939

\bibitem[{{Kausch} {et~al.}(2015){Kausch}, {Noll}, {Smette}, {Kimeswenger},
  {Barden}, {Szyszka}, {Jones}, {Sana}, {Horst}, \&
  {Kerber}}]{2015A&A...576A..78K}
{Kausch}, W., {Noll}, S., {Smette}, A., {et~al.} 2015, \aap, 576, A78

\bibitem[{{Kerzendorf} {et~al.}(2017){Kerzendorf}, {McCully}, {Taubenberger},
  {Jerkstrand}, {Seitenzahl}, {Ruiter}, {Spyromilio}, {Long}, \&
  {Fransson}}]{2017MNRAS.472.2534K}
{Kerzendorf}, W.~E., {McCully}, C., {Taubenberger}, S., {et~al.} 2017, \mnras,
  472, 2534

\bibitem[{{Krisciunas} {et~al.}(2003){Krisciunas}, {Suntzeff}, {Candia},
  {Arenas}, {Espinoza}, {Gonzalez}, {Gonzalez}, {H{\"o}flich}, {Landolt},
  {Phillips}, \& {Pizarro}}]{2003AJ....125..166K}
{Krisciunas}, K., {Suntzeff}, N.~B., {Candia}, P., {et~al.} 2003, \aj, 125, 166

\bibitem[{{Leloudas} {et~al.}(2009){Leloudas}, {Stritzinger}, {Sollerman},
  {Burns}, {Kozma}, {Krisciunas}, {Maund}, {Milne}, {Filippenko}, {Fransson},
  {Ganeshalingam}, {Hamuy}, {Li}, {Phillips}, {Schmidt}, {Skottfelt},
  {Taubenberger}, {Boldt}, {Fynbo}, {Gonzalez}, {Salvo}, \&
  {Thomas-Osip}}]{2009A&A...505..265L}
{Leloudas}, G., {Stritzinger}, M.~D., {Sollerman}, J., {et~al.} 2009, \aap,
  505, 265

\bibitem[{{Li} {et~al.}(2001){Li}, {Filippenko}, {Gates}, {Chornock},
  {Gal-Yam}, {Ofek}, {Leonard}, {Modjaz}, {Rich}, {Riess}, \&
  {Treffers}}]{2001PASP..113.1178L}
{Li}, W., {Filippenko}, A.~V., {Gates}, E., {et~al.} 2001, \pasp, 113, 1178

\bibitem[{{Li} {et~al.}(2019){Li}, {Wang}, {Hu}, {Yang}, {Zhang}, {Mo}, {Chen},
  {Zhang}, {Benetti}, {Cappellaro}, {Elias-Rosa}, {Isern}, {Morales-Garoffolo},
  {Huang}, {Ochner}, {Pastorello}, {Reguitti}, {Tartaglia}, {Terreran},
  {Tomasella}, \& {Wang}}]{2019ApJ...882...30L}
{Li}, W., {Wang}, X., {Hu}, M., {et~al.} 2019, \apj, 882, 30

\bibitem[{{Maguire} {et~al.}(2016){Maguire}, {Taubenberger}, {Sullivan}, \&
  {Mazzali}}]{2016MNRAS.457.3254M}
{Maguire}, K., {Taubenberger}, S., {Sullivan}, M., \& {Mazzali}, P.~A. 2016,
  \mnras, 457, 3254

\bibitem[{{Maguire} {et~al.}(2018){Maguire}, {Sim}, {Shingles}, {Spyromilio},
  {Jerkstrand}, {Sullivan}, {Chen}, {Cartier}, {Dimitriadis}, {Frohmaier},
  {Galbany}, {Guti{\'e}rrez}, {Hosseinzadeh}, {Howell}, {Inserra}, {Rudy}, \&
  {Sollerman}}]{2018MNRAS.477.3567M}
{Maguire}, K., {Sim}, S.~A., {Shingles}, L., {et~al.} 2018, \mnras, 477, 3567

\bibitem[{{Marion} {et~al.}(2016){Marion}, {Brown}, {Vink{\'o}}, {Silverman},
  {Sand}, {Challis}, {Kirshner}, {Wheeler}, {Berlind}, {Brown}, {Calkins},
  {Camacho}, {Dhungana}, {Foley}, {Friedman}, {Graham}, {Howell}, {Hsiao},
  {Irwin}, {Jha}, {Kehoe}, {Macri}, {Maeda}, {Mandel}, {McCully}, {Pandya},
  {Rines}, {Wilhelmy}, \& {Zheng}}]{2016ApJ...820...92M}
{Marion}, G.~H., {Brown}, P.~J., {Vink{\'o}}, J., {et~al.} 2016, \apj, 820, 92

\bibitem[{{Matheson} {et~al.}(2012){Matheson}, {Joyce}, {Allen}, {Saha},
  {Silva}, {Wood-Vasey}, {Adams}, {Anderson}, {Beck}, {Bentz}, {Bershady},
  {Binkert}, {Butler}, {Camarata}, {Eigenbrot}, {Everett}, {Gallagher},
  {Garnavich}, {Glikman}, {Harbeck}, {Hargis}, {Herbst}, {Horch}, {Howell},
  {Jha}, {Kaczmarek}, {Knezek}, {Manne-Nicholas}, {Mathieu}, {Meixner},
  {Milliman}, {Power}, {Rajagopal}, {Reetz}, {Rhode}, {Schechtman-Rook},
  {Schwamb}, {Schweiker}, {Simmons}, {Simon}, {Summers}, {Young}, {Weyant},
  {Wilcots}, {Will}, \& {Williams}}]{2012ApJ...754...19M}
{Matheson}, T., {Joyce}, R.~R., {Allen}, L.~E., {et~al.} 2012, \apj, 754, 19

\bibitem[{{Mattila} {et~al.}(2008){Mattila}, {Meikle}, {Lundqvist},
  {Pastorello}, {Kotak}, {Eldridge}, {Smartt}, {Adamson}, {Gerardy}, {Rizzi},
  {Stephens}, \& {van Dyk}}]{2008MNRAS.389..141M}
{Mattila}, S., {Meikle}, W.~P.~S., {Lundqvist}, P., {et~al.} 2008, \mnras, 389,
  141

\bibitem[{{Mayya} {et~al.}(1998){Mayya}, {Puerari}, \&
  {Kuhn}}]{1998IAUC.6907....2M}
{Mayya}, Y.~D., {Puerari}, I., \& {Kuhn}, O. 1998, \iaucirc, 6907

\bibitem[{{Meikle}(2000)}]{2000MNRAS.314..782M}
{Meikle}, W.~P.~S. 2000, \mnras, 314, 782

\bibitem[{{Modigliani} {et~al.}(2010){Modigliani}, {Goldoni}, {Royer},
  {Haigron}, {Guglielmi}, {Fran{\c c}ois}, {Horrobin}, {Bristow}, {Vernet},
  {Moehler}, {Kerber}, {Ballester}, {Mason}, \&
  {Christensen}}]{2010SPIE.7737E..28M}
{Modigliani}, A., {Goldoni}, P., {Royer}, F., {et~al.} 2010, in \procspie, Vol.
  7737, Observatory Operations: Strategies, Processes, and Systems III, 773728

\bibitem[{{Nozawa} {et~al.}(2011){Nozawa}, {Maeda}, {Kozasa}, {Tanaka},
  {Nomoto}, \& {Umeda}}]{2011ApJ...736...45N}
{Nozawa}, T., {Maeda}, K., {Kozasa}, T., {et~al.} 2011, \apj, 736, 45

\bibitem[{{Nugent} {et~al.}(1995){Nugent}, {Phillips}, {Baron}, {Branch}, \&
  {Hauschildt}}]{1995ApJ...455L.147N}
{Nugent}, P., {Phillips}, M., {Baron}, E., {Branch}, D., \& {Hauschildt}, P.
  1995, \apjl, 455, L147

\bibitem[{{Pan} {et~al.}(2015){Pan}, {Foley}, {Kromer}, {Fox}, {Zheng},
  {Challis}, {Clubb}, {Filippenko}, {Folatelli}, {Graham}, {Hillebrandt},
  {Kirshner}, {Lee}, {Pakmor}, {Patat}, {Phillips}, {Pignata}, {R{\"o}pke},
  {Seitenzahl}, {Silverman}, {Simon}, {Sternberg}, {Stritzinger},
  {Taubenberger}, {Vinko}, \& {Wheeler}}]{2015MNRAS.452.4307P}
{Pan}, Y.-C., {Foley}, R.~J., {Kromer}, M., {et~al.} 2015, \mnras, 452, 4307

\bibitem[{{Pastorello} {et~al.}(2007){Pastorello}, {Mazzali}, {Pignata},
  {Benetti}, {Cappellaro}, {Filippenko}, {Li}, {Meikle}, {Arkharov}, {Blanc},
  {Bufano}, {Derekas}, {Dolci}, {Elias-Rosa}, {Foley}, {Ganeshalingam},
  {Harutyunyan}, {Kiss}, {Kotak}, {Larionov}, {Lucey}, {Napoleone},
  {Navasardyan}, {Patat}, {Rich}, {Ryder}, {Salvo}, {Schmidt}, {Stanishev},
  {Sz{\'e}kely}, {Taubenberger}, {Temporin}, {Turatto}, \&
  {Hillebrandt}}]{2007MNRAS.377.1531P}
{Pastorello}, A., {Mazzali}, P.~A., {Pignata}, G., {et~al.} 2007, \mnras, 377,
  1531

\bibitem[{{Patat}(2005)}]{2005MNRAS.357.1161P}
{Patat}, F. 2005, \mnras, 357, 1161

\bibitem[{{Patat} {et~al.}(2006){Patat}, {Benetti}, {Cappellaro}, \&
  {Turatto}}]{2006MNRAS.369.1949P}
{Patat}, F., {Benetti}, S., {Cappellaro}, E., \& {Turatto}, M. 2006, \mnras,
  369, 1949

\bibitem[{{Pinto} \& {Eastman}(2000)}]{2000ApJ...530..757P}
{Pinto}, P.~A., \& {Eastman}, R.~G. 2000, \apj, 530, 757

\bibitem[{{Quinet}(1996)}]{1996A&AS..116..573Q}
{Quinet}, P. 1996, \aaps, 116, 573

\bibitem[{{Quinet} {et~al.}(1996){Quinet}, {Le Dourneuf}, \&
  {Zeippen}}]{1996A&AS..120..361Q}
{Quinet}, P., {Le Dourneuf}, M., \& {Zeippen}, C.~J. 1996, \aaps, 120, 361

\bibitem[{{Rest} {et~al.}(2012){Rest}, {Sinnott}, \&
  {Welch}}]{2012PASA...29..466R}
{Rest}, A., {Sinnott}, B., \& {Welch}, D.~L. 2012, \pasa, 29, 466

\bibitem[{{Ryan} {et~al.}(2018){Ryan}, {Casertano}, \&
  {Pirzkal}}]{2018PASP..130c4501R}
{Ryan}, R.~E., J., {Casertano}, S., \& {Pirzkal}, N. 2018, \pasp, 130, 034501

\bibitem[{{Sand} {et~al.}(2016){Sand}, {Hsiao}, {Banerjee}, {Marion},
  {Diamond}, {Joshi}, {Parrent}, {Phillips}, {Stritzinger}, \&
  {Venkataraman}}]{2016ApJ...822L..16S}
{Sand}, D.~J., {Hsiao}, E.~Y., {Banerjee}, D.~P.~K., {et~al.} 2016, \apjl, 822,
  L16

\bibitem[{{Schlafly} \& {Finkbeiner}(2011)}]{2011ApJ...737..103S}
{Schlafly}, E.~F., \& {Finkbeiner}, D.~P. 2011, \apj, 737, 103

\bibitem[{{Shappee} {et~al.}(2017){Shappee}, {Stanek}, {Kochanek}, \&
  {Garnavich}}]{2017ApJ...841...48S}
{Shappee}, B.~J., {Stanek}, K.~Z., {Kochanek}, C.~S., \& {Garnavich}, P.~M.
  2017, \apj, 841, 48

\bibitem[{{Skrutskie} {et~al.}(2006){Skrutskie}, {Cutri}, {Stiening},
  {Weinberg}, {Schneider}, {Carpenter}, {Beichman}, {Capps}, {Chester},
  {Elias}, {Huchra}, {Liebert}, {Lonsdale}, {Monet}, {Price}, {Seitzer},
  {Jarrett}, {Kirkpatrick}, {Gizis}, {Howard}, {Evans}, {Fowler}, {Fullmer},
  {Hurt}, {Light}, {Kopan}, {Marsh}, {McCallon}, {Tam}, {Van Dyk}, \&
  {Wheelock}}]{2006AJ....131.1163S}
{Skrutskie}, M.~F., {Cutri}, R.~M., {Stiening}, R., {et~al.} 2006, \aj, 131,
  1163

\bibitem[{{Smette} {et~al.}(2015){Smette}, {Sana}, {Noll}, {Horst}, {Kausch},
  {Kimeswenger}, {Barden}, {Szyszka}, {Jones}, {Gallenne}, {Vinther},
  {Ballester}, \& {Taylor}}]{2015A&A...576A..77S}
{Smette}, A., {Sana}, H., {Noll}, S., {et~al.} 2015, \aap, 576, A77

\bibitem[{{Sollerman} {et~al.}(2004){Sollerman}, {Lindahl}, {Kozma}, {Challis},
  {Filippenko}, {Fransson}, {Garnavich}, {Leibundgut}, {Li}, {Lundqvist},
  {Milne}, {Spyromilio}, \& {Kirshner}}]{2004A&A...428..555S}
{Sollerman}, J., {Lindahl}, J., {Kozma}, C., {et~al.} 2004, \aap, 428, 555

\bibitem[{{Spyromilio} {et~al.}(2004){Spyromilio}, {Gilmozzi}, {Sollerman},
  {Leibundgut}, {Fransson}, \& {Cuby}}]{2004A&A...426..547S}
{Spyromilio}, J., {Gilmozzi}, R., {Sollerman}, J., {et~al.} 2004, \aap, 426,
  547

\bibitem[{{Stritzinger} \& {Sollerman}(2007)}]{2007A&A...470L...1S}
{Stritzinger}, M., \& {Sollerman}, J. 2007, \aap, 470, L1

\bibitem[{{Taubenberger} {et~al.}(2015){Taubenberger}, {Elias-Rosa},
  {Kerzendorf}, {Hachinger}, {Spyromilio}, {Fransson}, {Kromer}, {Ruiter},
  {Seitenzahl}, {Benetti}, {Cappellaro}, {Pastorello}, {Turatto}, \&
  {Marchetti}}]{2015MNRAS.448L..48T}
{Taubenberger}, S., {Elias-Rosa}, N., {Kerzendorf}, W.~E., {et~al.} 2015,
  \mnras, 448, L48

\bibitem[{{Troja} {et~al.}(2017){Troja}, {Piro}, {van Eerten}, {Wollaeger},
  {Im}, {Fox}, {Butler}, {Cenko}, {Sakamoto}, \& {Fryer}}]{2017Natur.551...71T}
{Troja}, E., {Piro}, L., {van Eerten}, H., {et~al.} 2017, \nat, 551, 71

\bibitem[{{Truran} {et~al.}(1967){Truran}, {Arnett}, \& {Cameron}}]{Truran1967}
{Truran}, J.~W., {Arnett}, W.~D., \& {Cameron}, A.~G.~W. 1967, Canadian Journal
  of Physics, 45, 2315

\bibitem[{{Tully} {et~al.}(2016){Tully}, {Courtois}, \&
  {Sorce}}]{2016AJ....152...50T}
{Tully}, R.~B., {Courtois}, H.~M., \& {Sorce}, J.~G. 2016, \aj, 152, 50

\bibitem[{{Tyndall} {et~al.}(2016){Tyndall}, {Ramsbottom}, {Ballance}, \&
  {Hibbert}}]{2016MNRAS.462.3350T}
{Tyndall}, N.~B., {Ramsbottom}, C.~A., {Ballance}, C.~P., \& {Hibbert}, A.
  2016, \mnras, 462, 3350

\bibitem[{{Valenti} {et~al.}(2016){Valenti}, {Howell}, {Stritzinger}, {Graham},
  {Hosseinzadeh}, {Arcavi}, {Bildsten}, {Jerkstrand}, {McCully}, {Pastorello},
  {Piro}, {Sand}, {Smartt}, {Terreran}, {Baltay}, {Benetti}, {Brown},
  {Filippenko}, {Fraser}, {Rabinowitz}, {Sullivan}, \&
  {Yuan}}]{2016MNRAS.459.3939V}
{Valenti}, S., {Howell}, D.~A., {Stritzinger}, M.~D., {et~al.} 2016, \mnras,
  459, 3939

\bibitem[{{Vernet} {et~al.}(2011){Vernet}, {Dekker}, {D'Odorico}, {Kaper},
  {Kjaergaard}, {Hammer}, {Randich}, {Zerbi}, {Groot}, {Hjorth}, {Guinouard},
  {Navarro}, {Adolfse}, {Albers}, {Amans}, {Andersen}, {Andersen}, {Binetruy},
  {Bristow}, {Castillo}, {Chemla}, {Christensen}, {Conconi}, {Conzelmann},
  {Dam}, {de Caprio}, {de Ugarte Postigo}, {Delabre}, {di Marcantonio},
  {Downing}, {Elswijk}, {Finger}, {Fischer}, {Flores}, {Fran{\c c}ois},
  {Goldoni}, {Guglielmi}, {Haigron}, {Hanenburg}, {Hendriks}, {Horrobin},
  {Horville}, {Jessen}, {Kerber}, {Kern}, {Kiekebusch}, {Kleszcz}, {Klougart},
  {Kragt}, {Larsen}, {Lizon}, {Lucuix}, {Mainieri}, {Manuputy}, {Martayan},
  {Mason}, {Mazzoleni}, {Michaelsen}, {Modigliani}, {Moehler}, {M{\o}ller},
  {Norup S{\o}rensen}, {N{\o}rregaard}, {P{\'e}roux}, {Patat}, {Pena}, {Pragt},
  {Reinero}, {Rigal}, {Riva}, {Roelfsema}, {Royer}, {Sacco}, {Santin},
  {Schoenmaker}, {Spano}, {Sweers}, {Ter Horst}, {Tintori}, {Tromp}, {van
  Dael}, {van der Vliet}, {Venema}, {Vidali}, {Vinther}, {Vola}, {Winters},
  {Wistisen}, {Wulterkens}, \& {Zacchei}}]{2011A&A...536A.105V}
{Vernet}, J., {Dekker}, H., {D'Odorico}, S., {et~al.} 2011, \aap, 536, A105

\bibitem[{{Williams} {et~al.}(2012){Williams}, {Borkowski}, {Reynolds},
  {Ghavamian}, {Blair}, {Long}, \& {Sankrit}}]{2012ApJ...755....3W}
{Williams}, B.~J., {Borkowski}, K.~J., {Reynolds}, S.~P., {et~al.} 2012, \apj,
  755, 3

\bibitem[{{Yang} {et~al.}(2018){Yang}, {Wang}, {Baade}, {Brown}, {Cikota},
  {Cracraft}, {H{\"o}flich}, {Maund}, {Patat}, {Sparks}, {Spyromilio},
  {Stevance}, {Wang}, \& {Wheeler}}]{2018ApJ...852...89Y}
{Yang}, Y., {Wang}, L., {Baade}, D., {et~al.} 2018, \apj, 852, 89

\bibitem[{{Yang} {et~al.}(2019){Yang}, {Hoeflich}, {Baade}, {Maund}, {Wang},
  {Brown}, {Stevance}, {Arcavi}, {Burke}, {Cikota}, {Clocchiatti}, {Gal-Yam},
  {Graham}, {Hiramatsu}, {Hosseinzadeh}, {Howell}, {Jha}, {McCully}, {Patat},
  {Sand}, {Schulze}, {Spyromilio}, {Valenti}, {Vinko}, {Wang}, {Wheeler},
  {Yaron}, \& {Zhang}}]{2019arXiv190310820Y}
{Yang}, Y., {Hoeflich}, P.~A., {Baade}, D., {et~al.} 2019, arXiv e-prints,
  arXiv:1903.10820

\end{thebibliography}
\end{document}